\documentclass[aps,prl,twocolumn,showpacs]{revtex4}
\usepackage{graphicx}
\newcommand{\ds}{ _{\downarrow}}
\newcommand{\us}{ _{\uparrow}}
\newcommand{\up}{\uparrow}
\newcommand{\down}{\downarrow}

\begin{document}
\draft \title{Analytical theory of the dressed bound state in highly polarized Fermi gases}
\author{R. Combescot, S. Giraud and X. Leyronas} \address{Laboratoire de
Physique Statistique, Ecole Normale Sup\'erieure, 24 rue Lhomond,
75231 Paris Cedex 05, France}
\date{Received \today}

\begin{abstract}
We present an analytical treatment of a single $\down$ atom within a Fermi sea of $\up$ atoms, when the interaction
is strong enough to produce a bound state, dressed by the Fermi sea. Our method makes use of a diagrammatic
analysis, with the involved diagrams taking only into account at most two particle-hole pairs excitations.
The agreement with existing Monte-Carlo results is excellent.
In the BEC limit our equation reduces exactly to the Skorniakov and Ter-Martirosian equation.
We present results when $\up$ and $\down$ atoms have different masses, which is of interest for experiments in  progress.
\end{abstract}

\maketitle

The field of ultracold Fermi gases  \cite{gps} has turned recently toward very interesting
and unexplored physical domains for superfluidity. Indeed it is possible to have stable
mixtures of two fermionic species with different particle numbers \cite{rimit} (called polarized gases)
and also different masses. When quantum degeneracy is reached, these fermions may
form Cooper pairs, leading to BCS-like superfluidity. In the regime where these pairs
are very tightly bound, forming essentially small molecules, one obtains the Bose-Einstein
condensation (BEC) of these resulting bosons. However when the attractive interaction
gets weaker, which can be achieved at will experimentally through Feshbach resonances
\cite{gps}, these pairs are overlaping increasingly and one has to explore the whole extent
of the BEC-BCS crossover. Polarization has been shown to be detrimental to superfluidity
in this crossover,
but one could also hope to discover new superfluid phases, like the FFLO phases, although
experimental results are up to now negative in this respect.

A particularly attractive limiting case is the one of very strong polarization, which is at the
same time easier to handle but provides also the possibility to understand quantitatively
the physics at lower polarization \cite{lrgs,pg,rls}. This is the situation where
a single fermion
(say $\down$-atom) with mass $m\ds$ is in the presence of a Fermi sea of another, say $\up$, 
fermion species with mass $m \equiv m\us$, Fermi wave vector $k_F$ and scattering length $a$
between $\up$ and $\down$ atoms. 
In the absence of bound state, we have shown recently \cite{cg} that the exact solution of this problem is obtained 
as the limit of an extremely rapidly convergent series
of approximations. These successive approximations amount to restrict the Hilbert space of the excited states of the system to 
one, two,...,$n$,.. particle-hole pairs. In practice the first order approximation, which coincides with the standard ladder approximation
is already quite satisfactory. The second order one, where at most two particle-hole pairs are coming in, gives an essentially
exact answer, as it may be confirmed by comparison with QMC calculations, when available.

In this paper we address analytically, by a diagrammatic extension of the above treatment,
the case where the attractive interaction is strong enough to lead to a bound state,
which is merely a molecule in the 'BEC' limit of very strong attraction $k_Fa\rightarrow 0_+$.
This regime is of essential importance \cite{pg} for the understanding of the whole phase
diagram, since when the density of $\down$ atoms increases, the corresponding bound
states form a Bose-Einstein condensate.
We will calculate in this regime the chemical potential and the effective mass, with results
for the case of equal masses $m\us=m\ds$ in remarkable agreement with known results
from QMC calculations \cite{pg,ps}.
This allows us to provide accurate answers in the case of very high current interest where the 
mass ratio $r=m\ds /m$ is different from 1. In the BEC limit our results are exact, since they reduce to the
Skorniakov and Ter-Martirosian \cite{stm} equation.

It is physically quite clear \cite{gc} that the existence of a bound state corresponds to a pole in the vertex
corresponding to the forward scattering of the $\down$-atom and an $\up$-atom. Precisely one can
show that such a pole gives rise to a singularity in the self-energy $\Sigma\ds({\bf p},\omega )$
for $\omega <0$, leading to a non-zero density $n\ds \neq 0$ \cite{fw}. Since we have a single 
$\down$-atom $N\ds=1$, we are, in the thermodynamic limit, just at the border
between a zero density $n\ds = 0$ and a non-zero density $n\ds \neq 0$. 
Considering the evolution of the singularities of $\Sigma\ds({\bf p},\omega )$ when $\mu \ds$ 
is increasing, starting from very large negative values, we find
that, for $N\ds=1$, a singularity has just reached the border $\omega =0$, as in \cite{crlc}.
Moreover when the bound state just appears, this corresponds to a zero total energy $\Omega =0$
for the two atoms. Hence the energies of the two scattering atoms are zero in our case of interest.
The physical $\mu \ds$ is the lowest value either giving rise to a bound state or satisfying the condition
$\mu \ds=\Sigma({\bf 0},0)$, used in \cite{crlc}, when no bound state exists.
With respect to the momentum variables it is clear physically that the lowest energy for the pole is
obtained when the total momentum ${\bf q}$
of the two atoms is zero. More precisely the dependence of the chemical potential on this momentum
will give the effective mass, which is found positive in
the physical range.

Our approximation is the following: in all the diagrams we draw, we have at most two explicit propagator lines corresponding to $\up$-atoms
running forward through the diagram.
Naturally we have also the propagator line of our single $\down$-atom which runs forward
throughout the diagram. We could generalize to 3, 4 $\cdots$ explicit propagators, this series of approximations converging
very rapidly to the exact result, just as in \cite{cg}, but again the present approximation is quite enough.
On completely general grounds all the diagrams we can draw for the vertex begin with an
elementary interaction between our $\down$-atom and an $\up$-atom. However, since we let the strength $g$ of this interaction
go to zero $g \rightarrow 0$, we have to repeat this scattering an infinite number of times, any finite number giving a zero
contribution. The summation of this series gives a factor $T_2({\bf q},\Omega)$, instead of the factor $g$ we would have 
for a single interaction. It depends only on the total momentum ${\bf q}$ and energy $\Omega $ of the two scattering atoms, and not
on the variables of each atom separately. In vacuum we would have explicitly $T_2({\bf q},\Omega)=T^{(0)}_2({\bf q},\Omega)=
(2\pi /m_r)\left[a^{-1}-\sqrt{2m_r(q^2/2M-\Omega )}\right]^{-1}$ (i.e. essentially the scattering amplitude of the two atoms
with total mass $M$ and reduced mass $m_r$), 
but here we have to calculate this quantity in the presence of the Fermi sea, with chemical potentials $\mu \us$ and $\mu \ds$.
In a first level approximation, where we would have at most a single propagator line for $\up$-atoms, $T_2$ would be the only
contribution. However at our second level approximation, with two $\up$-atoms propagators, we have also to consider processes where, 
as a first interaction or after $T_2$, our
single $\down$-atom interacts with another $\up$-atom.
\begin{figure}
\centering
\includegraphics[width=80mm]{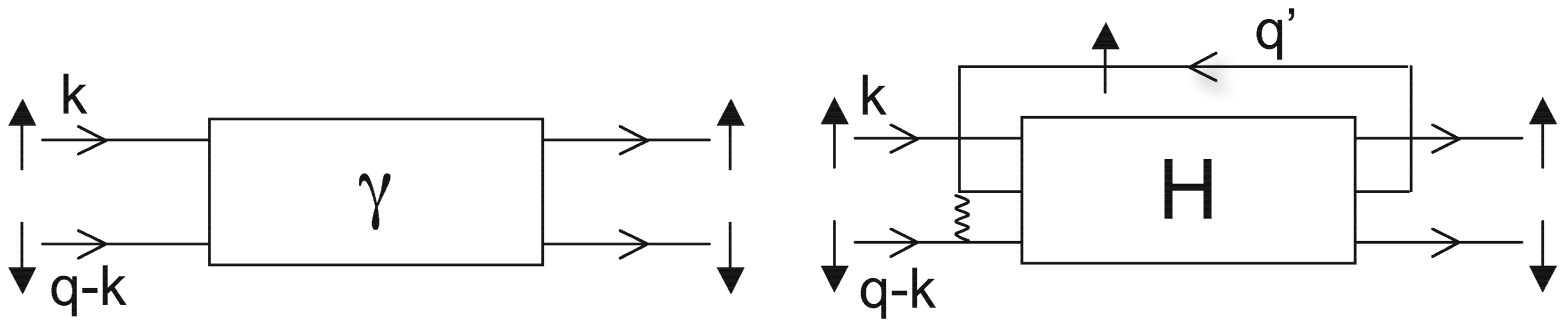}
\caption{Diagrammatic representation of $\gamma ({\bf k},{\bf q})$ and $H({\bf k},{\bf q},{\bf q}')$}
\vspace{-5mm}
\label{Figdiag}
\end{figure}

Before proceeding let us consider first what happens when we calculate within our approximation the self-energy $\Sigma\ds({\bf 0},0 )$
relevant for the case without bound state \cite{crlc}.
In this case the essential difference is that, in order to obtain $\Sigma$ from the vertex \cite{fw}, we should not set $\Omega =0$ but
rather close all the diagrams by a $\up$
propagator running backward. This gives an additional factor $G_{0 \up}({\bf q},\Omega)$ and we have then to sum over ${\bf q}$ and $\Omega $.
We sum over $\Omega $ by closing the $\Omega $ contour (which runs on the imaginary frequency axis)
at infinity in the half-plane ${\rm Re}\,\Omega<\,0$. We do similarly for the other $\up$ propagator
going backward. On the other hand, for all the explicit intermediate $\up$ propagators going forward
(that is in the same direction as our entering $\down$ propagator), we close the contour in the 
${\rm Re}\,\omega>\,0$ half-plane. It can then be seen that, due to the explicit 
$\up$ and $\down$ intermediate propagators, all the frequency integrations lead
merely to on-the-shell evaluations for the remaining factors. This is naturally so only when these propagators give rise
to poles in the corresponding frequency integration domain. Otherwise the result is zero. This leads
to the constraint $k>k_F$ on the wave vectors for the forward propagators, and $q<k_F$ for the ones
going backward. In this way we have been able to rederive exactly the equation ruling
the self-energy in \cite{cg} and accordingly all the results of \cite{cg} for $\mu \ds$.

Coming back to our present problem, the only difference is that, instead of having the on-the-shell
evaluation $\Omega = \epsilon _q - \mu \us$, with $\epsilon_{\bf q} ={\bf q}^{2}/2m$,
we have merely to set $\Omega =0$.
In the following we will have this implicit understanding for the variable named ${\bf q}$.
Otherwise, by proceeding in the same way as above, we will have on-the-shell evaluations
for all the other frequency variables entering our diagrams, with the same constraints
$k_i>k_F$ and $q_i<k_F$ on the momenta. Hence, since all the frequencies are
determined in this way, we refer in the following only to the momentum variables.

We need to write a general equation for the vertex $\gamma({\bf k},{\bf q})$
with entering momenta ${\bf k}$ for the $\up$-atom and
${\bf q}-{\bf k}$ for the $\down$-atom.
Except for a factor, this quantity is actually quite similar to $\alpha _{\bf k q}$ in \cite{cg},
the essential difference being the $\Omega=0$, instead of on-the-shell, value for the ${\bf q}$ variable. 
A first contribution is naturally $T_2({\bf q},0)$. However we have also the possibility 
that the first interaction of the $\down$-atom is with another $\up$-atom, with momentum ${\bf q}'$.
For fixed value of ${\bf q}'$, we call this contribution $H({\bf k},{\bf q},{\bf q}')$. 
Since ${\bf q}'$ is an internal
variable, one has to sum over it to obtain the contribution to $\gamma({\bf k},{\bf q})$. Finally, after
the repeated scattering described by $T_2({\bf q},0)$, we have the possibility to have the processes
described by H. These two parts will be linked by an $\up$ and a $\down$
propagator. Integration over the frequency of the $\up$ propagator gives a factor  $-1/{\bar E}^{(1)}_{{\bf k}{\bf q}}$ coming from the
on-the-shell evaluation of the $\down$ one. In this way we obtain:
\begin{eqnarray}\label{eqgam}
\gamma({\bf k},{\bf q})\!=\!T_2({\bf q},0)
\!\!\left[1+\!\sum_{{\bf k}'{\bf q}'}\frac{H({\bf k}',{\bf q},{\bf q}')}{{\bar E}^{(1)}_{{\bf k}'{\bf q}}}\right]
\!\!-\!\! \sum_{{\bf q}'}H({\bf k},{\bf q},{\bf q}')
\end{eqnarray}
with ${\bar E}^{(1)}_{{\bf k}{\bf q}}= |\mu \ds|+E_{{\bf k}-{\bf q}} + \epsilon _{\bf k}- \mu \us$ and 
$\,E_{\bf p}={\bf p}^2/2m\ds$,
quite analogous to $E^{(1)}_{{\bf k}{\bf q}}$ in \cite{cg}.

We want to find under which conditions $\gamma({\bf k},{\bf 0})$ diverges. This can naturally occur
if $[T_2({\bf 0},0)]^{-1}=0$, but this is just the first level approximation. However this can also arise from
a divergence of $H$, and we will look for the equation corresponding to this condition. Quite generally
we have for $\gamma$ and $H$ a set of coupled linear equations. The divergence is obtained
when there is a solution for the homogeneous part of these equations. In the following we retain only
the terms contributing to this homogeneous part and we omit the other ones.

We have now to write for $H({\bf k},{\bf q},{\bf q}')$ an equation analogous to Eq.(\ref{eqgam}). However, since we accept only two $\up$ propagators, we do not have to go to higher order vertices, and our equation will be closed.
First of all, the $\down$-atom and $\up$-atom, having a first interaction, will interact repeatedly,
giving rise to a common factor $T_2({\bf q}+{\bf q}'-{\bf k},\epsilon_{{\bf q}'}-\epsilon_{{\bf k}})$.
Then, after this process, we may have another $H$ vertex, giving rise to a term completely analogous  to the second
one in Eq.(\ref{eqgam}). However we have also the possibility that, after this repeated interaction,
the involved $\up$-atom recombines with a hole (this corresponds to closing the $T_2$ by a
backward ${\bf q}'$ propagator). Afterwards the ${\bf q}'$ $\up$-atom has disappeared
and all the possible remaining processes are described by $\gamma({\bf k},{\bf q})$. There is
again a factor $-1/{\bar E}^{(1)}_{{\bf k}{\bf q}}$ coming from the propagator linking these two parts.
Finally when we replace $\gamma({\bf k},{\bf q})$ by its explicit expression from Eq.(\ref{eqgam}),
the first term does not contain $H$ and we do not retain it. In this way, keeping only the homogeneous part of the equation, we obtain that the bound state appears when:
\begin{eqnarray*}\label{}
\left[T_2({\bf q}+{\bf q}'-{\bf k},\epsilon _{q'}-\epsilon _k)\right]^{-1}\!\!\!&\!\!\!\!\!H({\bf k},{\bf q},{\bf q}')=  \\ \nonumber
\sum_{{\bf k}'} \frac{H({\bf k}',{\bf q},{\bf q}')}{{\bar E}^{(2)}_{{\bf k}{\bf k}'{\bf q}{\bf q}'}}
-\frac{1}{{\bar E}^{(1)}_{{\bf k}{\bf q}}}&\sum_{{\bf q}''}H({\bf k},{\bf q},{\bf q}'')  \\ \nonumber
\end{eqnarray*}
\vspace{-10mm}
\begin{eqnarray}\label{}
\hspace{20mm}+\;\frac{T_2({\bf q},0)}{{\bar E}^{(1)}_{{\bf k}{\bf q}}} \sum_{{\bf k}'{\bf q}''}\frac{H({\bf k}',{\bf q},{\bf q}'')}{{\bar E}^{(1)}_{{\bf k}'{\bf q}}}
\end{eqnarray}
with ${\bar E}^{(2)}_{{\bf k}{\bf k}'{\bf q}{\bf q}'}= |\mu \ds|+E_{{\bf k}+{\bf k}'-{\bf q}-{\bf q}'} + \epsilon _{\bf k}+\epsilon _{\bf k'}-\epsilon _{\bf q'}- \mu \us$.
In this equation ${\bf q}$ appears as a parameter and again the lowest energy bound state is obtained
when ${\bf q}={\bf 0}$. Setting $H({\bf k},{\bf 0},{\bf q}') \equiv H_0({\bf k},{\bf q}')$ we find the following equation for the appearance
of the lowest energy bound state:
\begin{eqnarray}\label{eqbound}
\left[T_2({\bf q}-{\bf k},\epsilon _{q}-\epsilon _k)\right]^{-1}\;H_0({\bf k},{\bf q})=
\sum_{{\bf k}'} \frac{H_0({\bf k}',{\bf q})}{{\bar E}^{(2)}_{{\bf k}{\bf k}'{\bf 0}{\bf q}}} \\ \nonumber
-\frac{1}{{\bar E}^{(1)}_{{\bf k}{\bf 0}}}\sum_{{\bf q}'}H_0({\bf k},{\bf q}')
+\frac{T_2({\bf 0},0)}{{\bar E}^{(1)}_{{\bf k}{\bf 0}}} \sum_{{\bf k}'{\bf q}'}\frac{H_0({\bf k}',{\bf q}')}{{\bar E}^{(1)}_{{\bf k}'{\bf 0}}}
\end{eqnarray}
We show now analytically that, in the BEC limit, our equation leads systematically to the exact result, whatever the mass ratio $r$.

In this BEC limit our equation simplifies since it can be seen as the limit $k_F\rightarrow 0$ at fixed $a$. Since by definition $q \leq k_F$,
we are left with a function $H_0({\bf k},{\bf 0}) \equiv h({\bf k})$ of the single variable ${\bf k}$ (actually only of its modulus for symmetry reasons).
Then the summation over ${\bf q}'$ in the last two terms of Eq.(\ref{eqbound}) gives merely a factor $n\us=k_F^3/(6\pi ^2)$. On the other hand,
when we compare the first two terms in the right-hand side of Eq.(\ref{eqbound}), the only difference is that the typical range for the summation
over ${\bf k}'$ is $k' \sim 1/a$ (as it can be seen explicitly below) while it is $q' \sim k_F$ for the summation over ${\bf q}'$, as we have just seen.
Hence the second term is of order $(k_Fa)^3$ compared to the first, and accordingly completely negligible in the BEC limit (actually we find
numerically that, even for the lowest relevant values of $1/k_Fa$ this term gives a very small contribution). Accordingly
we are left with:
\begin{eqnarray}\label{eqbec}
\left[T_2({\bf k},\!-\epsilon _k)\right]^{-1}\!h({\bf k})\!=\! \sum_{{\bf k}'} \frac{h({\bf k}')}{{\bar E}^{(2)}_{{\bf k}{\bf k}'{\bf 0}{\bf 0}}}\!
+\!\frac{k_F^3}{6\pi ^2}\frac{T_2({\bf 0},0)}{{\bar E}^{(1)}_{{\bf k}{\bf 0}}} \sum_{{\bf k}'}\frac{h({\bf k}')}{{\bar E}^{(1)}_{{\bf k}'{\bf 0}}}
\end{eqnarray}
We are looking for a solution with the form $\rho \equiv |\mu \ds|/E_F=\epsilon _b/E_F+1+\beta (m/m_r)k_Fa$ which is an expansion
in powers of $k_Fa$, with $\epsilon _b=1/2m_r a^2$ the molecule binding energy and $E_F=k_F^2/2m$ the Fermi energy. 
Taking into account the general expression of $T_2({\bf k},\omega )$ which gives $\left[m_rT_2({\bf 0},0)/(2\pi a)\right]^{-1}
=1-(2k_Fa/\pi )\left[1+\sqrt{(\rho-1)m_r/m} \arctan (\sqrt{(\rho-1)m_r/m})\right]$ and substituting the expansion for $\rho$ gives the expansion
$\left[m_rT_2({\bf 0},0)/(2\pi a)\right]^{-1}=-(k_Fa)^3\left[\beta/2+2/(3\pi )\right]$. The lowest level approximation corresponds to
write $\left[T_2({\bf 0},0)\right]^{-1}=0$. This leads to $\beta=-4/(3\pi )$ which corresponds to the Born approximation for the $\up$-atom-dimer
scattering length.

While in the BEC limit we have, in the last term of Eq.(\ref{eqbec}), to evaluate
$T_2({\bf 0},0)$ carefully since it is large, all the other terms can be evaluated to lowest order. Hence we set $|\mu \ds|/E_F=(m/m_r)/(k_Fa)^2$
in ${\bar E}^{(1)}_{{\bf k}{\bf 0}}=|\mu \ds|+k^2/2m_r$ and in ${\bar E}^{(2)}_{{\bf k}{\bf k}'{\bf 0}{\bf 0}}=|\mu \ds|+({\bf k}+{\bf k}')^2/2m\ds+(k^2+k'^2)/2m$.
Similarly we have $\left[m_rT_2({\bf k},-\epsilon _k)/(2\pi a)\right]^{-1}=1-\sqrt{1+R(ka)^2}$, where $R=m_r/m_r^T$ is the ratio of $m_r$
to the reduced mass $m_r^T=m(m+m\ds)/(2m+m\ds)$ of the $\up$-atom-dimer system ($R=3/4$ for equal masses). We take the reduced variable
${\bf x}=a{\bf k}$ and make the change of function $h({\bf k})=C\,f(x)/x^2$ where $C$ is a constant to be determined just below. We find from Eq.(\ref{eqbec}):
\begin{eqnarray}\label{eqstm}
\hspace{-25mm}\pi R\,\!\left[1+\sqrt{1+Rx^2}\right]^{-1} \,f(x)&=& \\ \nonumber
\end{eqnarray}
\vspace{-10mm}
\begin{eqnarray*}\label{}
-\frac{1}{2\pi} \int_{0}^{\infty} dy \int d\Omega_y \frac{f(y)}{1+x^2+y^2+R'\,{\bf x}.{\bf y}} \\ \nonumber
+\frac{8}{4+3\pi \beta}\frac{1}{1+x^2} \int_{0}^{\infty}dy\,\frac{f(y)}{1+y^2}&&
\end{eqnarray*}
where $R'=2m_r/m\ds$ and the angular average over ${\bf y}$ can be easily performed. 
We can choose $C$ so that $f(0)=-3 \pi \beta/(2R)$. Then, writing Eq.(\ref{eqstm}) for $x=0$, we find
that the coefficient of $1/(1+x^2)$ (including the integral) in the last term of Eq.(\ref{eqstm}) is just equal to $\pi $. Hence Eq.(\ref{eqstm})
coincides exactly with the equation, suitably generalized to the case of unequal masses, found by Skorniakov and Ter-Martirosian 
\cite{stm} for the scattering amplitude to obtain the fermion-dimer scattering length $a_3$. Solving Eq.(\ref{eqstm}) provides
the exact result for $a_3= a f(0)$. From the above relation this gives $\beta=-2R a_3/(3\pi a)$. When this is inserted in the above
expression for $|\mu \ds|$, we obtain a contribution $-2\pi n\us a_3/m_r^T$ from the fermion-dimer scattering, which is precisely
the one resulting from a mean-field argument, exact in this limit. Hence our equation provides the exact answer in the BEC limit by reducing to the
Skorniakov and Ter-Martirosian equation.

\begin{figure}
\centering
\includegraphics[width=80mm]{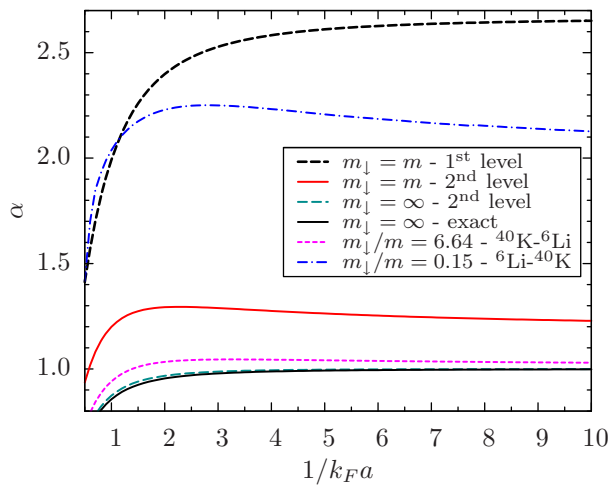}
\caption{Reduced chemical potential $\alpha  \equiv(3\pi m_r^T/2mE_F k_Fa)$ $(\epsilon _b+E_F-|\mu \ds|)$ as a function of $1/k_Fa$
for various mass ratios.}
\vspace{-5mm}
\label{figmu}
\end{figure}

We have solved numerically Eq.(\ref{eqbound}) in the general case. The results for $\mu \ds$ are given in Fig.~\ref{figmu}, or rather we plot
$\alpha  \equiv(3\pi m_r^T/2mE_F k_Fa)$ $(\epsilon _b+E_F-|\mu \ds|)$ which reduces to $a_3/a$ in the BEC limit and allows to magnify small differences.
For equal masses $m\ds=m$ we find that, basically down to $1/k_Fa \sim 2$, $\alpha $ is essentially constant, almost equal to its BEC value $a_3/a=1.18$
(we display also the first level result $[T_2({\bf 0},0)]^{-1}=0$ which goes to the Born result $a_3/a=8/3$ in the BEC limit).
This is exactly what is found by QMC calculations \cite{pg,ps}.
More precisely we find the actual value of $\alpha $ is slightly higher, the decrease toward the BEC limit being quite slow, behaving as $k_Fa$.
We also find that the
bound state appears for $1/k_Fa \simeq 0.88$ in perfect agreement with QMC \cite{ps} (all the more
remarkable since the two chemical potential curves for the polaron and the bound state cross at a very small
angle, which makes this value quite sensitive). Another very striking check of the precision of our results is found for $m\ds = \infty$, where the exact result is known \cite{crlc}. For most of the range the agreement
is within a few $10^{-3}$ and the difference is barely seen even in our blown up Fig.~\ref{figmu}. A very interesting feature of this limit is that the convergence toward the BEC result $\alpha =1$ is as $(k_Fa)^2$, faster
than in the general case. This allows to understand qualitatively why, for the equal mass case, which is not much different, one obtains also a fairly
slow variation. The case of physical interest $r=6.64$ for the $^{40}$K-$^6$Li mixture is also shown and is actually very close to this limit.
We display also the $r=1/6.64 \simeq 0.15$ result. Note finally that, in agreement with \cite{cg}, the quite simple ${\bf q}={\bf 0}$ approximation of Eq.(\ref{eqbound})
gives results in fairly good agreement with our exact numerical treatment.

\begin{figure}
\centering
\includegraphics[width=80mm]{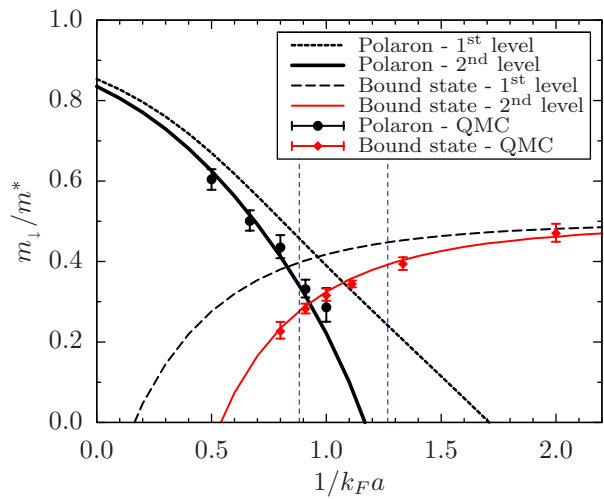}
\caption{Reduced inverse effective mass $m\ds/m* $ of the polaron and of the bound state as a function of $1/k_Fa$, at the first
and second level approximations. The dashed lines indicate the location of the appearance of the bound
state, both at the first level $1/k_Fa \simeq 1.27$ and the second level $1/k_Fa \simeq 0.88$ approximations.}
\vspace{-5mm}
\label{figmasseff}
\end{figure}

Finally we show in Fig.~\ref{figmasseff} our results for the effective mass for equal masses $m\ds=m$.
The agreement with diagrammatic QMC results \cite{ps} is again quite remarkable, both on the polaron side
and on the bound state side. It is noteworthy that the effective mass is essentially continuous at the transition.
It is tempting to speculate that this is an exact result, which could be checked experimentally. This is physically
reasonable since, at the threshold for bound state appearance, the polaron and the bound state are physically identical objects. 
This physical fact is supported by the $m\ds = \infty$ explicit case. Note also
that when the effective mass is calculated beyond this transition $m\ds/m* $ becomes negative, as seen in
Fig.~\ref{figmasseff}, which signals an instability. As expected physically the transition between polaron and
bound state occurs before the occurrence of this instability. For $m\ds = \infty$ the bound state appears naturally
for $1/k_Fa=0$. On the other hand when one gets to lighter $m\ds$, the transition value for $1/k_Fa$ goes toward
higher and higher values. This is already seen easily at our first level approximation which, although not accurate
as seen in Fig.~\ref{figmasseff}, are nevertheless qualitatively correct.

In conclusion we have shown that our equations provide an essentially exact analytical description of the molecular state dressed by a Fermi sea.

The ``Laboratoire de Physique Statistique'' is ``Laboratoire associ\'e au Centre National de la Recherche
Scientifique et aux Universit\'es Paris 6 et Paris 7''.

\end{document}